\documentclass[sigconf]{acmart}
\usepackage{multirow} 
\usepackage{subfig}
\usepackage{wrapfig}
\usepackage{enumitem}
\usepackage[ruled]{algorithm2e}
\usepackage{algorithmic}
\usepackage{amsmath}
\usepackage{bm}
\usepackage{tabu}                      
\usepackage{booktabs}                  
\usepackage{lipsum}                    
\usepackage{mwe}                       
\usepackage{amsmath}

\AtBeginDocument{%
  }

\copyrightyear{2025}
\acmYear{2025}
\setcopyright{rightsretained}
\acmConference[CHI EA '25]{Extended Abstracts of the CHI Conference on Human Factors in Computing Systems}{April 26-May 1, 2025}{Yokohama, Japan}
\acmBooktitle{Extended Abstracts of the CHI Conference on Human Factors in Computing Systems (CHI EA '25), April 26-May 1, 2025, Yokohama, Japan}\acmDOI{10.1145/3706599.3719908}
\acmISBN{979-8-4007-1395-8/2025/04}

\begin{document}

\title[Optimizing Curve-Based Selection with On-Body Surfaces in Virtual Environments]{Optimizing Curve-Based Selection with On-Body Surfaces in Virtual Environments}


\author{Xiang Li}
\orcid{0000-0001-5529-071X}
\affiliation{%
  \institution{University of Cambridge}
  \city{Cambridge}
  \country{United Kingdom}
}
\email{xl529@cam.ac.uk}

\author{Per Ola Kristensson}
\orcid{0000-0002-7139-871X}
\affiliation{%
  \institution{University of Cambridge}
  \city{Cambridge}
  \country{United Kingdom}
}
\email{pok21@cam.ac.uk}

\renewcommand{\shortauthors}{Xiang Li and Per Ola Kristensson}

\begin{abstract}

Virtual Reality (VR) interfaces often rely on linear ray-casting for object selection but struggle with precision in dense or occluded environments. This late-breaking work introduces an optimized dual-layered selection mechanism combining dynamic Bézier Curves, controlled via finger gestures, with on-body interaction surfaces to enhance precision and immersion. Bézier Curves offer fine-grained control and flexibility in complex scenarios, while on-body surfaces project nearby virtual objects onto the user’s forearm, leveraging proprioception and tactile feedback. A preliminary qualitative study ($N$ = 24) compared two interaction paradigms (Bézier Curve vs. Linear Ray) and two interaction media (On-body vs. Mid-air). Participants praised the Bézier Curve’s ability to target occluded objects but noted the physical demand. On-body interactions were favored for their immersive qualities, while mid-air interactions were appreciated for maintaining focus on the virtual scene. These findings highlight the importance of balancing ease of learning and precise control when designing VR selection techniques, opening avenues for further exploration of curve-based and on-body interactions in dense virtual environments.

\end{abstract}




\begin{teaserfigure}
  \centering
  \includegraphics[width=\linewidth]{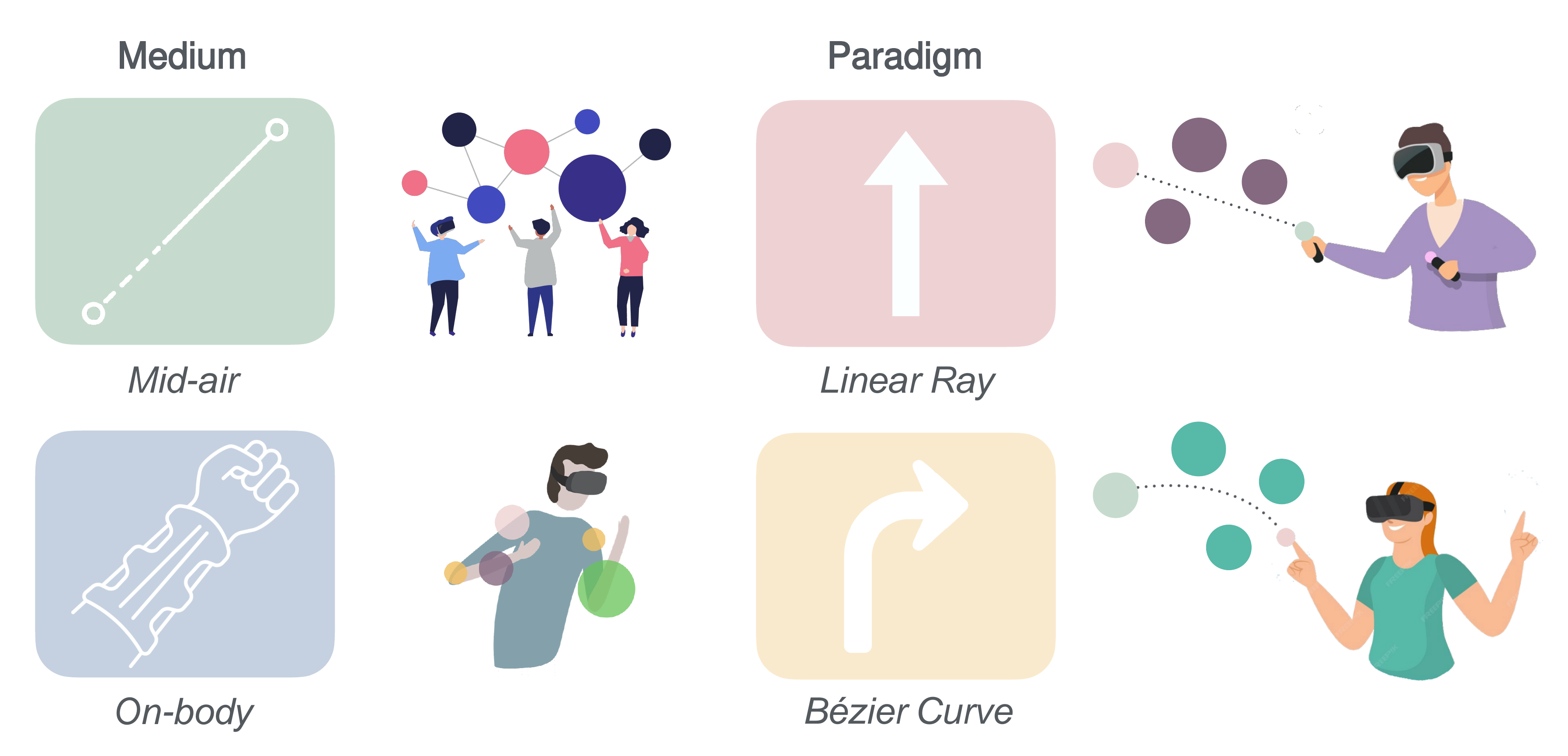}
  \caption{We explore four selection techniques for selection tasks in VR by examining two interaction media (On-body and Mid-air) and two interaction paradigms (Linear Ray and Bézier Curve).}
  \Description{We provided four selection techniques for selection tasks in VR by examining two interaction media (On-body and Mid-air) and two interaction paradigms (Linear Ray and Bézier Curve).}
  \label{fig:teaser}
\end{teaserfigure}

\maketitle

\section{Introduction}

Virtual Reality (VR) has transitioned from a niche concept to a versatile platform for gaming, training, and remote collaboration. As VR environments grow increasingly complex, designing efficient interaction techniques becomes essential to enhance user engagement and task performance. Traditional interaction methods, such as mid-air gestures and linear ray-casting, remain prevalent due to their simplicity~\cite{bowman_evaluation_1997,jr_3d_2017,pierce_voodoo_1999}. However, in dense or cluttered environments, these methods often result in reduced accuracy, increased physical or cognitive demand, and user frustration. To address these limitations, alternative approaches have explored leveraging interaction media and rethinking interaction paradigms. On-body surfaces, which use the user’s body as an interaction medium, present an innovative solution~\cite{yu_blending_2022}. These surfaces provide passive haptic feedback and leverage proprioceptive cues, which are our innate sense of body position and movement, to enhance spatial awareness and reduce cognitive load~\cite{simeone2014feet}. For example, tapping or gesturing on the arm or leg can improve precision and engagement in selection tasks~\cite{gustafson_imaginary_2011, harrison_skinput_2010,yu_blending_2022}.

Another promising direction is introducing curve-based interaction paradigms, such as Bézier Curves, which allow users to interact with occluded objects by dynamically adjusting selection paths. Unlike linear ray-casting, curve-based techniques are more flexible and effective in resolving occlusions, reducing the need for physical repositioning~\cite{olwal_flexible_2003,steinicke_object_2004}. Additionally, the ergonomic benefits of curve-based selection can simplify gestures and reduce user fatigue, particularly in high-density environments~\cite{bohm1984survey,lu_investigating_2020}.

In this late-breaking work, we present a novel interaction system that integrates Bézier Curve-based selection with on-body surfaces to address the challenges of object selection in dense VR environments. Our system dynamically generates Bézier Curves based on the real-time curvature of the user’s fingers, enabling precise selection of occluded targets. To complement this, we introduce a proximity matching projection mechanism that maps the nearest virtual objects to the user's forearm, improving the accuracy and immersion of the selection.

A preliminary user study with 24 participants provided valuable insights into the user experience of our techniques. Participants identified challenges with the precision of the proximity-matching mechanism, particularly under occlusion or positional variability, emphasizing the need for refinement. On the other hand, participants appreciated the immersive qualities of on-body interactions and the precision of Bézier Curves in dense environments but noted physical demands during extended use. These findings highlight the potential of combining curve-based selection techniques with on-body surfaces to create more efficient, immersive, and user-friendly VR interfaces. Our contributions can be summarized as follows:

\begin{itemize}
    \item We propose an algorithm that generates Bézier Curves using the real-time curvature of the user’s fingers.
    \item We design and optimize a proximity-matching projection mechanism for on-body surfaces, seamlessly integrated with selection techniques.
    \item We report the results of a preliminary user study where we gathered qualitative feedback on the curve-based selection technique for on-body surfaces in virtual environments, identifying key user experience themes.
\end{itemize}

\section{Related Work}
\subsection{Body-Centric Interaction}

On-body interaction techniques use the human body as an interactive platform, significantly enhancing immersion and proprioception in VR environments~\cite{harrison_-body_2012,bergstrom_human--computer_2019,coyle_i_2012,floyd_mueller_limited_2021,mueller_towards_2023,li2021vrcaptcha}. These approaches improve accuracy in eyes-free targeting, owing to the natural familiarity of users with their own body dimensions \cite{weigel_skinmarks_2017,gustafson_imaginary_2010}. Extensive research has been conducted on utilizing various body parts, such as arms, palms---even the skin—as interfaces for VR interactions \cite{chatain_digiglo_2020,dezfuli_palmrc_2014,mistry_wuw_2009}. Innovations like Skinput \cite{harrison_skinput_2010} and Touché \cite{sato_touche_2012} demonstrate advanced gesture recognition capabilities by detecting acoustic and capacitive signals directly from the skin.

Further studies have assessed the effectiveness of interfaces anchored to the non-dominant arm for precise pointing tasks in VR, as explored by Li et al. \cite{li_armstrong_2021}. The development of body-centric selection and manipulation techniques such as the Hand Range Interface \cite{xu_hand_2018}, Swarm Manipulation \cite{li2023swarm,li2024swarm}, BodyLoci \cite{fruchard_impact_2018}, and BodyOn \cite{yu_blending_2022} has opened new possibilities for enhancing mid-air interactions using the human body itself as an interface. Additionally, on-body menus such as the Tap-tap Menu \cite{azai_tap-tap_2018} and PalmGesture \cite{wang_palmgesture_2015} have been pivotal in exploring how visual and tactile cues can be effectively utilized to navigate these innovative user interfaces \cite{li2024onbodymenu}. These investigations highlight the significant potential of on-body interaction to create more engaging VR experiences.

\subsection{Mid-Air Interaction}

Mid-air interaction is a prominent feature in modern headset-based VR systems, allowing users to interact with digital content in virtual environments through gestures and movements. These interactions are often mediated by game controllers or directly through hand movements \cite{cornelio_martinez_agency_2017,koutsabasis_empirical_2019,song_hotgestures_2023}. Renowned for its straightforward approach, mid-air interaction is particularly adept in 3D spaces due to its versatile input capabilities \cite{jr_3d_2017}. Users need first to hover the hand over the virtual object and then perform a grab gesture to select the object \cite{xu2019dmove}. However, it is also criticized for its lack of precision \cite{arora_experimental_2017,mendes2019survey,argelaguet_survey_2013}, the potential for user fatigue \cite{xu2020exploring,hinckley_survey_1994}, and the absence of tactile feedback \cite{fang_retargeted_2021}, which can diminish the overall user experience.

In response to these challenges, researchers have explored methods to improve the usability and broaden the interaction vocabulary of mid-air systems. Studies have considered more relaxed input methods, such as adopting an arms-down posture to reduce fatigue \cite{brasier2020arpads,liu2015gunslinger}. In addition, computational techniques have been employed to improve input precision, for example, by developing models that optimize the selection distribution \cite{yu_modeling_2019,yu_modeling_2023}. These advancements aim to mitigate the limitations of mid-air interactions while leveraging their inherent benefits in immersive environments.

\subsection{Selection Techniques for VR}

The ray-casting technique is commonly used for object selection and interaction in VR. This method, often described as pointing, involves projecting a ray from a source to select objects that intersect with it, similar to using a laser pointer \cite{hinckley_survey_1994}. Although initial implementations used handheld ray emitters with 5 Degrees of Freedom (DOF) for a clear selection method, now prevalent in both research and commercial VR devices, other adaptations include head-mounted rays utilizing head or gaze movements for orientation, typically offering 2 DOF \cite{hinckley_survey_1994}. Various enhancements, such as gaze-directed rays \cite{forsberg_aperture_1996}, and aperture circles \cite{forsberg_aperture_1996}, along with specific hand poses for interaction \cite{pierce_image_1997}, have been developed to improve selection accuracy and flexibility.

However, these methods often face challenges, such as hand tremors which can reduce precision, leading to the selection of multiple or unintended targets \cite{olsen_laser_2001, grossman_design_2006, yu_fully-occluded_2020}. To address these issues, predictive algorithms and spatial-temporal models have been employed \cite{schjerlund_ninja_2021, olwal_flexible_2003, olwal_senseshapes_2003, steed_towards_2006, von_willich_densingqueen_2023}, though the lack of visual feedback can complicate user understanding. Additionally, strategies such as snap targets are used to aid in selection in the midst of tremors, although their effectiveness remains uncertain \cite{forsberg_aperture_1996}. Furthermore, complications arise when rays intersect multiple targets; various techniques such as the Shadow Cone for hand angle adjustments \cite{steed_3d_nodate} and Depth Ray for depth-controlled selection \cite{grossman_design_2006} have been developed to mitigate these issues. Advanced interaction techniques like the Go-Go technique allow users to interact with distant objects by virtually extending their arms \cite{poupyrev_go-go_1996}, and two-step processes involve selecting a group of targets followed by using secondary tools for precise selection \cite{azai_tap-tap_2018, lediaeva_evaluation_2020}.

Despite these advancements, comparative studies indicate varied performance of ray-casting and other techniques under different experimental conditions \cite{kopper_rapid_2011, bowman_testbed_1999, looser_evaluation_2007, poupyrev_egocentric_1998, vanacken_exploring_2007}. While ray-casting is particularly effective for interacting with distant targets, it can be cumbersome for nearby objects \cite{cashion_dense_2012}. Further research into enhancements such as the bubble mechanism could potentially improve the usability and efficiency of ray-casting techniques in VR \cite{cashion_dense_2012}.

\section{Technical Design}

In this section, we outline the technical design of our interactive system, specifically focusing on (1) the generation of Bézier Curves from user gestures and (2) a proximity-matching mechanism optimized for real-time calculation. We explore the transformation of user gesture data into Bézier Curve parameters and the real-time projection of objects near the curve onto a predetermined on-body surface (i.e., the forearm).

\subsection{Bézier Curve Formulation}

We chose the Bézier Curve for curve fitting because it has been widely used in computer graphics and has proven ergonomic benefits that simplify interaction~\cite{olwal_flexible_2003,lu_investigating_2020,de_haan_intenselect_nodate,steinicke_object_2004}. A quadratic Bézier Curve, integral to our design, is defined by two endpoints (i.e., start point and end point) and a single control point: $\bm{P}_{\text{0}}$, $\bm{P}_{\text{1}}$, and $\bm{P}_{\text{2}}$~\cite{bohm1984survey, Post1989CurvesAS}. The mathematical expression of the curve is:
\begin{equation}
    \bm{B}(t) = (1 - t)^2 \bm{P}_{\text{0}} + 2(1 - t)t \bm{P}_{\text{1}} + t^2 \bm{P}_{\text{2}}, \quad (0 \leq t \leq 1)
\end{equation}
where \( t \) is a parameter that determines the position along the curve, with \( t = 0 \) at the start point and \( t = 1 \) at the end point.

\subsection{Gesture to Bézier Mapping}

In our system, the curvature of a virtual ray is controlled by the flexion of the user’s left index finger. This flexion dynamically adjusts the Bézier Curve parameters, allowing the system to modulate the ray’s curvature in real time. Specifically, the length between the fingertip and the wrist is measured to determine the degree of finger flexion. This distance modulates the curvature parameter $\kappa$, which directly influences the shape of the Bézier Curve.

\begin{figure*}[ht]
    \centering
    \includegraphics[width=\linewidth]{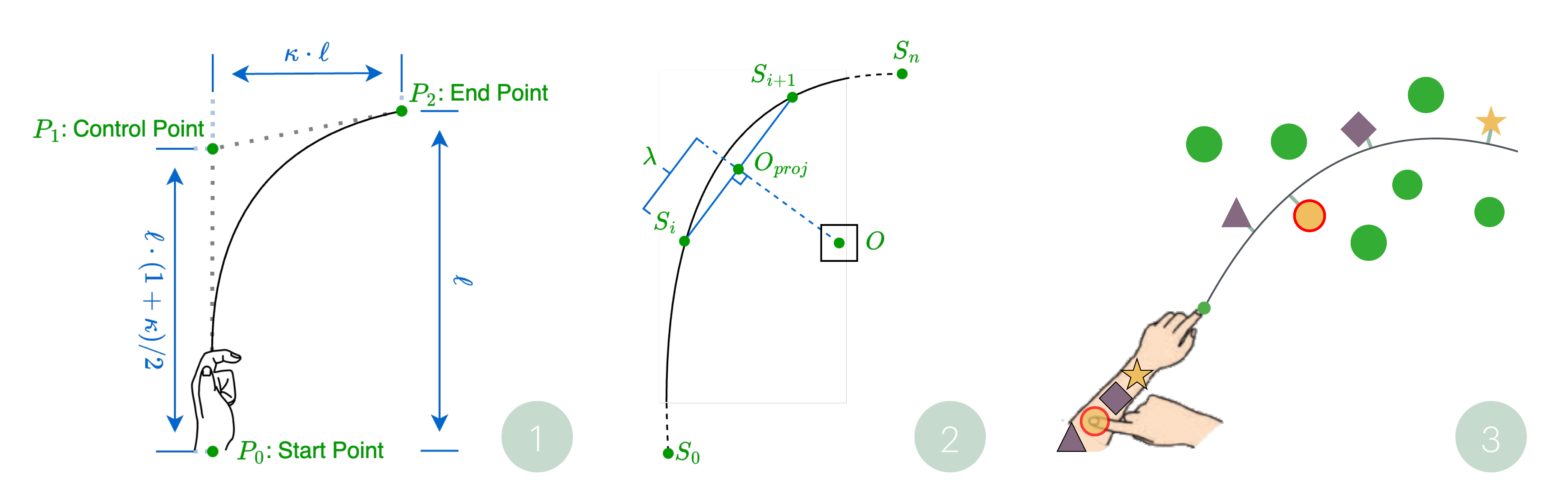}
    \caption{(1) Generation of a Bézier Curve defined by start point \( P_0 \), control point \( P_1 \), and end point \( P_2 \), using parameters such as curvature \( \kappa \) and length \( \ell \). (2) Example of how the proximity-matching mechanism works, demonstrating the calculation of the shortest distance from an object \( O \) to points on the curve, specifically between \( S_i \) and \( S_{i+1} \), and the projection \( O_{proj} \). (3) A schematic representation of using the on-body Bézier Curve with the proximity-matching mechanism, where the closest four objects are selected and projected onto the user's forearm.}
    \Description{A combined figure illustrating: (1) Bézier Curve generation with control and endpoint parameters, (2) proximity-matching mechanism with object projection, and (3) on-body interaction using Bézier Curves for virtual object selection.}
    \label{fig:merged_system}
\end{figure*}

The gesture capture process begins by defining the initial positions of the wrist and fingertip (denoted as $\bm{P}_0$ and $\bm{H}_1$, respectively) when the finger is fully extended (see Figure~\ref{fig:merged_system} (1)). These positions are expressed in 3D Cartesian coordinates $(x, y, z)$, where $x$, $y$, and $z$ correspond to the horizontal, vertical, and depth dimensions. The maximum distance, $L_{\text{straight}}$, is calculated using the Euclidean distance formula:
\begin{equation}
    L_{\text{straight}} = \sqrt{(x_1-x_0)^2 + (y_1-y_0)^2 + (z_1-z_0)^2}.
\end{equation}

As the finger flexes, this distance decreases to $L_{\text{bent}}$, and the curvature, $\kappa$, is computed as:
\begin{equation}
    \kappa = K_1 \cdot \left(\frac{L_{\text{straight}} - L_{\text{bent}}}{L_{\text{straight}}}\right)
\end{equation}
where $K_1 = 1.5$ is empirically determined to optimize interaction learnability and proved effective for the authors; however, this coefficient can be adjusted flexibly.

The initial point \( \bm{P}_{\text{0}} \) is positioned at the wrist, correlating with the virtual left hand's juncture. We establish \( \bm{v}_{\text{align}} \) as the unit vector extending parallel to the longitudinal axis of the forearm and \( \bm{v}_{\text{ortho}} \) as the unit vector perpendicular to the hand's plane. These vectors are instrumental in the determination of the locations for \( \bm{P}_{\text{1}} \) and \( \bm{P}_{\text{2}} \), which are modulated by \( \kappa \) and \( \ell \) as follows:

\begin{align}
    \bm{P}_{\text{1}} &= \bm{P}_{\text{0}} + \bm{v}_{\text{align}} \cdot \frac{1}{2} \cdot (1 + \kappa) \cdot \ell, \\
    \bm{P}_{\text{2}} &= \bm{P}_{\text{0}} + \ell \cdot \bm{v}_{\text{align}} + \kappa \cdot \ell \cdot \bm{v}_{\text{ortho}}.
\end{align}

Through the modulation of \( \kappa \) by the user's gesture of flexing their left index finger, the curvature of the Bézier ray is directly influenced. The dynamic modulation of the endpoint \( \bm{P}_{\text{2}} \) guarantees its alignment parallel to the wrist, thus facilitating an easy gestural interaction. Moreover, the calculated position of the control point \( \bm{P}_{\text{1}} \), derived from \( \kappa \), allows for the direct manipulation of the flexing trajectory of the curve, thereby synchronizing the Bézier Curve with the movements and curvature of the hand. Therefore, we have our quadratic Bézier Curve, which is defined by \( \bm{P}_{\text{0}} \), \( \bm{P}_{\text{1}} \), and \( \bm{P}_{\text{2}} \).

\subsection{Proximity-Matching Mechanism}

We initially revisited the challenges associated with using pointing gestures in VR~\cite{yu_fully-occluded_2020}. Subsequently, we opted for on-body menu designs~\cite{lediaeva_evaluation_2020,azai_tap-tap_2018,li2024onbodymenu}, which organize objects into a floating panel encircling the user's body, despite necessitating an additional step for selection. To optimize the utilization of limited space for projecting objects onto the user's forearm, we developed a proximity-matching mechanism, which can synchronously calculate the closest four objects to the curve and project them on the forearm.

Our system computes the minimum Euclidean distance from a point $\bm{O}$ to a discretized Bézier Curve for efficient real-time interaction. The Bézier Curve is approximated using 20 linear segments to streamline computational processes. Each segment is defined by points on the curve:
\begin{equation}
    \bm{S}_i = \bm{B}\left(\frac{i}{n}\right), \quad i \in \{0, 1, 2, ..., n\}
\end{equation}
where $n = 20$, and $\bm{B}(t)$ denotes the curve parametrized by $t$.

The minimum distance, $d_{\text{min}}$, from the point $\bm{O}$ to the curve is computed as the smallest distance to any segment:
\begin{equation}
    d_{\text{min}} = \min_{i} \left( d(\bm{O}, [\bm{S}_i, \bm{S}_{i+1}]) \right)
\end{equation}
where $d(\bm{O}, [\bm{S}_i, \bm{S}_{i+1}])$ measures the Euclidean distance from $\bm{O}$ to the linear segment between $\bm{S}_i$ and $\bm{S}_{i+1}$.

The orthogonal projection $\bm{O}_{\text{proj}}$ of $\bm{O}$ onto each segment is critical for determining interaction points accurately and swiftly:
\begin{equation}
    \bm{O}_{\text{proj}} = \bm{S}_i + \lambda (\bm{S}_{i+1} - \bm{S}_i)
\end{equation}
where $\lambda$ is calculated as:
\begin{equation}
    \lambda = \left( \frac{(\bm{O} - \bm{S}_i) \cdot (\bm{S}_{i+1} - \bm{S}_i)}{\| \bm{S}_{i+1} - \bm{S}_i \|^2} \right)
\end{equation}

The computation of $\lambda$ is further optimized using Unity's built-in functionality:
\begin{equation}
    \lambda(\bm{S}_{i+1} - \bm{S}_i) = \textit{Vector3.Project}(\bm{O} - \bm{S}_i, \bm{S}_{i+1} - \bm{S}_i)
\end{equation}

This methodology leverages Unity's efficient vector operations to ensure that the real-time computation of $\lambda$ is both accurate and responsive, which is essential for maintaining the fluidity and precision of user interactions in our system (see Figure~\ref{fig:merged_system} (3)).

Finally, user selection activation and locking mechanisms are controlled through hand gestures as well. Activation occurs when the right middle finger is bent, which initiates the appearance of the ray and enables the dynamic mapping of objects onto the left forearm in real time. Conversely, straightening the right middle finger locks the selection, causing the ray to disappear and stabilizing the mapping of objects on the left forearm. This facilitates a more straightforward selection of the target object by the user.

\section{Preliminary Qualitative User Study}

To investigate the performance and usability of on-body surfaces and assess the potential of our proposed Bézier Curve-based techniques, we conducted a preliminary user study with 24 participants (9 females, 15 males) recruited from a local university. Participants ranged in age from 18 to 29 years ($M = 22.46, SD = 1.60$). Some participants reported prior VR experience ($M = 4.14, SD = 2.35$), with familiarity ratings on a 7-point Likert scale (1 = no experience, 7 = expert). All participants were right-handed and used Meta Quest 2 headsets during the study. Drawing upon the differences between Interaction \textsc{Medium} and Interaction \textsc{Paradigm}, we used four selection techniques in our study: (1) mid-air Linear Ray, (2) mid-air Bézier Curve, (3) on-body Linear Ray, and (4) on-body Bézier Curve. 

\paragraph{Mid-air Linear Ray} This serves as our baseline condition, mirroring the default setting of the Meta Quest. Participants were instructed to use their dominant hand to point at the target and then confirm the selection with the other hand.

\paragraph{Mid-air Bézier Curve} Similar to the mid-air Linear Ray, but our Bézier Curve-generating algorithm replaces the Linear Ray. Participants still used their right hand to confirm the selection when the curve aligned with the target.

\paragraph{On-body Linear Ray} In contrast to mid-air interaction, participants observed the four projected objects on their forearms. After confirming that the target object had been successfully projected onto their forearm, they extended the index finger of their dominant hand to rigidly display the target and then touched the target to finalize the selection.

\paragraph{On-body Bézier Curve} Similar to the on-body Linear Ray, this condition incorporated the Bézier Curve-generating algorithm instead of the Linear Ray. \newline

Our proximity-matching mechanism was implemented specifically for the on-body conditions. To mitigate the issue of hand tremors, we used a bimanual interaction pattern, where participants used their left hand to indicate the target and their right hand to confirm the selection. The right hand remained in a closed fist when the selection was pending. When the user extended their index finger, this action was detected by the VR headset and interpreted as user confirmation.

\subsection{Procedure}
A 5-minute training session was carried out before each experimental condition, allowing participants to acclimate to the respective interaction technique. At the beginning of each round, 64 objects were randomly generated within an invisible cube with a side length of 1.5~m and a width of 3~m positioned 2.5~m in front of the participant, with one of the objects randomly highlighted to indicate the target. The objects were composed of icons of various colors and shapes, aiming to maximize user recognition and minimize bias induced by different familiarity levels of different users. The overall study comprised a total of 2,880 trials (2 (interaction media) $\times$ 2 (interaction paradigm) $\times$ 30 (repeat) $\times$ 24 (participant)).

\subsection{Preliminary Results from Interviews}
We analyzed the interview data using an inductive thematic analysis approach~\cite{braun2006using} to articulate two user experience themes: challenges with matching mechanisms and usability of interaction modalities.

\paragraph{Challenges with matching mechanisms}  
Participants reported challenges with the precision and responsiveness of the proximity-matching mechanism, particularly under occlusion or positional variability. For example, several participants noted, \textit{``sometimes it cannot be selected when occluded''} (P2), \textit{``the selection often depends on the user’s position''} (P5), and \textit{``the ray is not very responsive and is sometimes blocked''} (P8). Another participant highlighted that \textit{``when my finger and the camera are parallel, the mechanism sometimes fails to register properly''} (P4). These comments point to the need for improved handling of occlusions, better alignment, and enhanced responsiveness.

\paragraph{Usability of interaction modalities}  
Participants shared varied feedback on the interaction techniques. The Bézier Curve was praised for its precision in dense environments but was criticized for its physical demands. As P2 stated, \textit{``the curved one is not as good as the straight one; it cannot bend to the degree I want and feels less responsive.''} Similarly, P6 mentioned, \textit{``the Bézier Curve required too much finger flexion, which became tiring during extended use.''} In contrast, the Linear Ray was valued for its simplicity and natural gestures, but one participant noted, \textit{``the ray worked well for simple tasks but struggled with occluded targets or dense objects''} (P7). On-body interactions were appreciated for their immersive qualities, as one participant shared, \textit{``the physical contact gave a strong sense of control and confirmation''} (P10). However, limitations included \textit{``the need for high recognition accuracy with both hands,''} which occasionally caused errors during confirmation (P3).

\section{Discussion and Conclusion}
This late-breaking work introduces an optimized curve-based selection technique combined with on-body interaction surfaces to address the challenges of object selection in VR. Bézier Curves enable precise targeting of occluded objects, while on-body interactions enhance proprioception and haptic feedback, improving both accuracy and immersion. Preliminary findings revealed key trade-offs: Bézier Curves were praised for their precision in dense environments but induced fatigue due to finger flexion. Some participants found them ``intuitive'' after practice, while others preferred the simplicity of Linear Rays, despite their limitations in occlusion-heavy scenarios. On-body interactions were valued for their accuracy and tactile engagement, whereas mid-air gestures while maintaining visual focus on the scene, were perceived as less engaging because of the lack of haptic feedback.

Our findings emphasize the importance of balancing ease of learning and precise control in VR interaction design. On-body surfaces provided immersive and proprioceptive benefits, while Linear Rays and mid-air gestures offered simpler interactions in less complex environments. However, we acknowledge that the proximity-matching mechanism introduced approximation errors due to Bézier Curve discretization, which will be refined in future iterations.

Furthermore, this late-breaking work focuses primarily on system design and lacks an extensive quantitative evaluation. Future work will include systematic user studies to assess performance across varying object sizes, spatial densities, and scene configurations, incorporating quantitative metrics such as task completion time, error rates, and workload perception. We also plan to improve the accuracy of proximity-matching mechanisms by refining adaptive discretization and distance calculations, thereby optimizing interaction efficiency for diverse VR environments.

\begin{acks}
Xiang Li is supported by the China Scholarship Council (CSC) International Cambridge Scholarship (No. 202208320092). Per Ola Kristensson is supported by the EPSRC (grant EP/W02456X/1). 
\end{acks}

\bibliographystyle{ACM-Reference-Format}
\bibliography{bezier}

\end{document}